# Water Production by Comet 103P/Hartley 2 Observed with the SWAN Instrument on the SOHO Spacecraft

M.R. Combi[1*], J.-L. Bertaux[2], E. Quémerais[2], S. Ferron[3], J.T.T. Mäkinen[4]

[1]Dept. of Atmospheric, Oceanic and Space Sciences
University of Michigan
2455 Hayward Street
Ann Arbor, MI 48109-2143 USA
*Corresponding author: mcombi@umich.edu

[2]LATMOS, CNRS/INSU,
Université de Versailles Saint-Quentin
11 Bd d'Alembert, 78280 Guyancourt FRANCE

[3]ACRI-st, Sophia-Antipolis, FRANCE

[4]Finnish Meteorological Institute, Box 503
SF-00101 Helsinki, FINLAND

M.R. Combi: mcombi@umich.edu
J.L. Bertaux: Jean-Loup.Bertaux@aerov.jussieu.fr
E. Quémerais: eric.quemerais@latmos.ipsl.fr
S. Ferron: Stephane.Ferron@aerov.jussieu.fr
J.T.T. Mäkinen: Teemu.Makinen@fmi.fi

ABSTRACT

Global water production rates were determined from the Lyman-α emission of hydrogen around comet 103P/Hartley 2, observed with the SWAN (Solar Wind ANisotropies) all-sky camera on the SOHO spacecraft from September 14 through December 12, 2010. This time period included the November 4 flyby by the EPOXI spacecraft. Water production was 3 times lower than during the 1997 apparition also measured by SWAN. In 2010 it increased by a factor of ~2.5 within one day on September 30 with a similar corresponding drop between November 24 and 30. The total surface area of sublimating water within ±20 days of perihelion was ~0.5 km$^2$, about half of the mean cross section of the nucleus. Outside this period it was ~0.2 km$^2$. The peak value was 90%, implying a significant water production by released nucleus icy fragments.

1. INTRODUCTION

Comets are among the most volatile and least processed remnants of the nebula out of which our solar system was formed 4.5 Gyr ago. The appearances of short-period, so-called Jupiter-family comets (JFC) in the inner solar system are understood to originate from gravitational perturbations of icy objects in the Kuiper belt outside the orbit of Neptune in a process whereby they are passed down from Neptune to Uranus, then to Saturn, and finally to Jupiter. From there many can be sent into orbits that have perihelia in the vicinity of 1 AU (Fernandez 1980; Levison & Duncan 1997) and observable from the Earth. Typical JFCs have periods in the range of 6 to 8 years and are then seen to be active from the Earth for a few months around perihelion.

Comet 103P/Hartley 2 is one such JFC with an orbital period of 6.5 years and a perihelion distance of 1.05 AU. It was discovered on June 4, 1986, by Malcolm Hartley at the Siding Spring Observatory (Hartley 1986). It was seen during the 1991, 1997 and 2004 apparitions, though in 2004 it remained near superior conjunction on the other side of the sun during most of its active perihelion phase. Observations from 1991 and 1997 indicated a maximum water production rate of $\sim 3 \times 10^{28}$ molecules $s^{-1}$ but a sharp drop with increasing heliocentric distance (A'Hearn et al. 1995; Crovisier et al. 1999; Colangeli et al. 1999; Fink 2009; Combi et al. 2011). Infrared observations with the Spitzer Space Telescope (Lisse et al. 2009) made when the comet was not far past its most recent aphelion gave an estimate of its mean radius to be rather small (0.57 km) implying the nucleus had to have a mostly active water sublimating surface. 103P/Hartley 2 was ultimately chosen for a flyby during the extended EPOXI mission for the Deep Impact spacecraft (A'Hearn et al. 2011), which had impacted and made important measurements of JFC 9P/Tempel 1 in July 2005 (A'Hearn & Combi 2007).

Here we report the results of the analysis of observations of the hydrogen coma of comet 103P/Hartley 2 observed with the all-sky SWAN H Lyman-α camera on the SOlar and Heliospheric Observatory (SOHO) spacecraft made during 3 months around the 2010 apparition including the time period of the EPOXI flyby. From these we monitor the total global water production rate of the comet as a function of time, which also provides important activity context for other observations.

## 2. SOHO/SWAN OBSERVATIONS

The SOHO spacecraft has been in a halo orbit around the Earth-Sun L1 Lagrange point observing the Sun and solar wind since its launch in late 1995. The Solar Wind ANisotropies instrument is an all-sky scanning imager operating at the wavelength of neutral H Lyman-α at 121.6 nm (Bertaux et al. 1995). The main purpose of SWAN is to measure the Lyman-α emission of the interstellar neutral hydrogen that streams through the solar system and is carved out by the outflowing solar wind providing a global picture of anisotropies in the solar wind flow.

Atomic hydrogen is the most abundant species in the atmosphere (or coma) of nearly all comets. Most hydrogen is produced in a photodissociation chain originating with water molecules and the OH radicals produced from water (Combi & Smyth 1988; Crovisier 1989). Water is understood to be the most abundant volatile species in most comet nuclei and is believed to control the activity of the coma when comets are within ~3 AU from the sun. Because of their large hydrogen comae comets are easily observed by SWAN.

SWAN has observed the H Lyman-α coma of many comets (e.g., Bertaux et al., 1999, Mäkinen et al. 2001, Combi et al. 2005 & 2011). Measurements of the abundance and distribution of hydrogen in the coma can provide reliable estimates of water production rates in comets (Mäkinen & Combi 2005, Combi et al. 2005 & 2011, Feldman et al. 2004). Observations of 103P/Hartley 2 were made with the standard pipeline mode of daily full-sky observations.

SWAN has two sensor units, SU+Z and SU-Z, that typically observe north and south of the ecliptic, respectively. The current sensitivity of SU+Z is such that an intensity of 1 Rayleigh results in 0.24 counts per second per pixel. SU-Z is less sensitive than SU+Z by a factor of 2.6. Each sensor has an instantaneous field of view (IFOV) of 5°x5° in a multi-anode detector of 25 1°x1° pixels. Images are made by mosaicking the IFOV across the sky in 2.5° increments. Because of a spacecraft roll maneuver on October 29, 2010, all the observations of 103P/Hartley 2 were made with the SU+Z sensor even though the comet moved south of the ecliptic.

## 3. 103P/HARTLEY 2 OBSERVATIONS AND ANALYSIS

The full-sky SWAN images were examined beginning on August 1, 2010 for comet 103P/Hartley 2. The first image with a firm detection of the comet was the image on September 14, 2010. Thereafter, the comet was detected daily until October 16, when the comet was obstructed by the SOHO spacecraft itself. The comet reemerged unobstructed on October 29 and was observed on most days until December 12, 2010. On a number of days the comet was too close to field stars, especially during the period from November 24 to 30, to enable a clean signal from the comet to be isolated. Images of the position the comet on November 4 and November 10, 2010 are shown in Figure 1a-1b.

We used the model analysis procedure described for the SWAN observations of comet 1996 B2/Hyakutake (Mäkinen & Combi 2005). It combines the methods behind the syndyname (Keller & Meier 1976) and the vectorial models (Festou 1981), while considering coma-wide variations of input parameters and incorporating the necessary physical phenomena through the inclusion of a parameterized (and less computationally intensive) version of the H atom velocity distribution from Monte Carlo simulations (Combi & Smyth 1988) that account for the expansion of the coma and partial thermalization of escaping H atoms.

Water production rates were calculated for each usable SWAN image from September 14 to December 12, 2010. The dissociation chain of water to OH radicals and the H atoms produced, plus their transit times to fill the observable coma, introduces a time delay from any change in water activity near the nucleus to an observable coma response of 1 to 2 days. The large SWAN IFOV furthermore smears the significant short-term periodic variation produced by the rotation of the nucleus (A'Hearn et al. 2011). The close geocentric distance of the comet (0.11 to 0.35 AU) helps minimize this effect compared with previous some SWAN observations of comets (Mäkinen & Combi 2005). Table 1 gives the observational circumstances as well as the resulting water production for the SWAN observations. The g-factor is calculated from the composite solar Lyman-alpha data taken from the LASP web site http://lasp.colorado.edu/lisird/lya and the solar Lyman-$\alpha$ line profile by P. Lemaire et al. (1998). Figure 1c-d shows the H coma as observed by SWAN a few hours after the EPOXI flyby, on November 4, as well as a brightness profile cut through the coma.

Figure 2 shows the variation of the water production rate as a function time with the date of the EPOXI flyby noted. The water production rate increased slowly from 1.5 to 2.5 x $10^{27}$ molecules s$^{-1}$ from September 14 to 29 but then increased to 6.0 x $10^{27}$ molecules s$^{-1}$ in 1 day. Approaching perihelion it continued increasing until October 16 to a value of 8.7 x $10^{27}$ molecules s$^{-1}$ at which time the location of the comet in the sky became obstructed by the SOHO spacecraft. Then 3.7 days after perihelion the comet reemerged. The water production rate then varied rather irregularly between 5.5 x $10^{27}$ and 1.2 x $10^{28}$ molecules $s^{-1}$ until November 24 when it was difficult to locate the comet among field stars again until November 30. By this time the water production rate had dropped to a level slightly higher than the pre-September-30 level but lower than the mean perihelion ±20-day level. The mean water production rate determined from the SWAN observation on the day of the EPOXI flyby was 8.5 x $10^{27}$ molecules $s^{-1}$.

## 4. RESULTS AND DISCUSSION

The water production rate variation from September 14 through September 24 shows no evidence of correlation with the reported outburst of CN activity from the EPOXI team (A'Hearn et al. 2011), but seems more similar to the variation of their dust-scattered continuum observations, which increase rather monotonically throughout this period.

SWAN observed 103P/Hartley 2 during its 1997 apparition (Combi et al. 2011) yielding water production rates that were consistent with values determined from both ground-based observations of OH (A'Hearn et al. 1995) in 1991, $O(^1D)$ atoms (Fink 2009) in 1997 and from observations with the ISOPHOT instrument on the Infrared Space Observatory (Crovisier et al. 1999; Colangeli et al. 1999). SWAN results from 1997 and 2010 are compared in Figure 2. Clearly the comet and its production of water changed dramatically from 1997 (and 1991) to 2010. The 1997 production rates were a factor of 3 larger than those in the 2010 apparition.

One way to characterize water production rates in comets is to calculate an equivalent surface area of water ice, which when exposed to sunlight at the comet's heliocentric distance, is required to produce the observed water vapor. Because of the reality of variable surface and surface fractional coverage by water this is called the "minimum active area," It was calculated for all SWAN water production rates of 103P/Hartley 2 from 1997 and 2010 and compared with the measured minimum, maximum and mean cross sections of the nucleus from EPOXI imaging (A'Hearn et al. 2011), and all are plotted in Figure 3. The minimum active area is similar but not equal to the active area. It is defined as $A = LQr^2/[N_A F_S(1-A_V)]$, where L=50 kJ mol$^{-1}$ is the latent heat of water for sublimation, r is the heliocentric distance in AU, $N_A$=6.022 x 10$^{23}$ mol$^{-1}$ (the Avogadro constant), FS=1365 W m$^{-2}$ (the solar constant), and $A_V$=0.03 (the assumed bond albedo of the nucleus). See (Keller 1990) for a discussion of this definition.

For the majority of JFCs, the minimum active area calculated from the water production rate is typically between 5 and 20% of the physical surface area of their nuclei. Such was the case for the previous spacecraft flyby target comets: 1P/Halley, 19P/Borrelly, 81P/Wild 2 and 9P/Tempel 2 (A'Hearn et al. 1995; Fink 2009; Keller et al. 1987; Soderblom et al. 2002; Brownlee et al. 2006). 1P/Halley, which is not a JFC, had the largest active fraction of these at about 20%. The active area results for Hartley 2 are in many ways similar to those for long-period Oort cloud comet 1996 B2/Hyakutake in that the minimum active area has been comparable to or even larger than the projected cross section of the nucleus itself ($\pi R_N^2 \sim 1$ million meters$^2$). This was more so the case during the 1997 apparition of 103P/Hartley 2 when the minimum active area was more than 3 times the mean projected cross section of the nucleus. During outbursts of comet Hyakutake (Combi et al. 2005), the total production rate increased by a factor of 4 above the 'normal' level, when many fragments were released from the nucleus, including some large ones that were seen traveling down the tail for many days and producing an extended source of gas (Harris et al. 1997; Desvoivres et al. 2000).

The EPOXI results (A'Hearn et al. 2011) show that the activity of the surface of 103P/Hartley 2 is not distributed uniformly over most of its surface. So even during the 2010 apparition, the fact that the minimum active area peaks at the value of the mean projected area of the nucleus indicates that a significant fraction of its water production results from the extended halo of icy fragments that appear to be carried off the surface by $CO_2$-driven activity. The fact that the activity was three times larger in 1997 means either that this process was far more prevalent in 1997 (and also in 1991) or that some drastic alteration has occurred to the nucleus since 1997, or both.

The synthesis of all the observations (EPOXI, space-based, and ground-based) of this comet over the coming months and years will hopefully shed some light on a number of fundamentally important issues of cometary science. Model analysis of observations of water and its byproducts having a higher spatial resolution than the SWAN observations might be able to quantitatively separate water production directly from the nucleus and that from the extended cloud of fragments seen in the EPOXI images (A'Hearn et al. 2011) and shown to dominate the global rate measured by SWAN. The role of highly volatile species (e.g., $CO_2$) may have been underestimated or at least underappreciated in the activity of comets, perhaps even inside the solar-system's so-called snow line, challenging the current paradigms for cometary activity.

## ACKNOWLEDGEMENTS

SOHO is an international cooperative mission between ESA and NASA. M. Combi acknowledges support from grant NNG08AO44G from the NASA Planetary Astronomy Program. J. -L. Bertaux, E. Quémerais and S. Ferron acknowledge support from CNRS and CNES. J.T.T. Mäkinen was supported by the Finnish Meteorological Institute. We thank Dr. M.F. A'Hearn for helpful discussions of preliminary EPOXI results prior to publication.

## REFERENCES

A'Hearn, M. F., Millis, R. L., Schleicher, D. G., Osip, D. J., & Birch, P. V. 1995, Icarus, 118, 223

A'Hearn, M.F., & Combi, M.R. 2007, Icarus, 187, 1

A'Hearn, M.F. et al. 2011, Science, submitted

Bertaux, J.-L. et al. 1995, Solar Phys., 162, 403

Bertaux, J.-L., Kyrölä, E., Quemerais, E., Lallement, R., Schmidt, W., Summanen, T., Costa, J., & Mäkinen, T. 1999, Space Sci. Rev., 87, 129

Brownlee, D. et al. 2006, Science, 314, 1711

Colangeli, L. et al. 1999, A&A, 343, L87

Combi, M.R., Lee, Y., Pattel, T.S., Mäkinen, J.T.T., Bertaux, J.-L., & Quemérais, E. 2011, AJ, in press

Combi, M.R., Mäkinen, J.T.T., Bertaux, J.-L., & Quemérais, E. 2005, Icarus, 177, 228

Combi, M.R., & Smyth, W.H. 1988, ApJ, 327, 1044

Crovisier, J. 1994, A&A, 213, 459

Crovisier, J. et al. 1999, ESA SP- 427, 161

Desvoivres, E. et al. 2000, Icarus, 144, 172

Fernández, J. 1980, MNRAS, 192, 481

Feldman, P.D., Cochran, A.L., & Combi, M.R. 2004, in Comets II, M.C. Festou, H.U. Keller, H.A. Weaver, Eds., Univ. Arizona Press, pp. 425-447

Festou, M.C. 1981, A&A, 95, 69

Fink, U. 2009, Icarus, 301, 311

Hartley, M., 1986, IAU Circ. 4197, 1, Green, D.W.E., Ed.

Harris, W.M. et al. 1997, Science, 277, 676

Keller, H.U. 1990, in Physics and Chemistry of Comets, W.F. Huebner, Ed, Springer-Verlag, Berlin, pp. 18-21

Keller, H.U. et al. 1987, A&A, 187, 807

Keller, H.U., Meier, R.R. 1976, A&A, 52, 272

Lemaire, P., Emerich, C., Curdt, W., Schuehle, U., & Wilhelm, K. 1998, A&A, 334, 1095

Levison, H., & Duncan, M. 1997, Icarus, 127, 13

Lisse, C.M. et al. 2009, PASP, 121, 968

Mäkinen, J.T.T., Bertaux, J.-L., Combi, M.R., & Quémerais, E. 2001, Science, 292, 1326

Mäkinen, J.T.T., & Combi, M.R. 2005, Icarus, 177, 217

Soderblom, L. et al. 2002, Science, 296, 1087

Table 1. SOHO/SWAN Observations of Comet 103P/Hartley 2 and Water Production Rates in 2010

| $\Delta T$ (days) | r (AU) | $\Delta$ (AU) | g (s$^{-1}$) | $Q \pm \delta Q$ ($10^{27}$ s$^{-1}$) |
|---|---|---|---|---|
| -42.853 | 1.211 | 0.285 | 0.002205 | 1.81 ± 1.0 |
| -41.853 | 1.204 | 0.278 | 0.002192 | 2.29 ± 0.73 |
| -40.853 | 1.198 | 0.271 | 0.002192 | 2.42 ± 0.65 |
| -39.766 | 1.191 | 0.264 | 0.002192 | 2.07 ± 0.73 |
| -38.766 | 1.185 | 0.257 | 0.002191 | 1.91 ± 0.74 |
| -36.766 | 1.173 | 0.243 | 0.002191 | 3.56 ± 0.53 |
| -34.766 | 1.162 | 0.230 | 0.002179 | 2.47 ± 0.55 |
| -33.766 | 1.156 | 0.224 | 0.002178 | 2.37 ± 0.60 |
| -32.765 | 1.151 | 0.217 | 0.002178 | 2.76 ± 0.73 |
| -31.766 | 1.146 | 0.211 | 0.002167 | 2.39 ± 0.66 |
| -30.765 | 1.141 | 0.205 | 0.002167 | 3.04 ± 0.75 |
| -29.766 | 1.136 | 0.199 | 0.002167 | 2.05 ± 0.96 |
| -28.765 | 1.131 | 0.193 | 0.002167 | 2.91 ± 0.77 |
| -27.765 | 1.126 | 0.187 | 0.002167 | 2.52 ± 0.81 |
| -26.765 | 1.122 | 0.181 | 0.002156 | 6.17 ± 0.42 |
| -25.765 | 1.117 | 0.176 | 0.002156 | 6.07 ± 0.47 |
| -24.740 | 1.113 | 0.170 | 0.002155 | 6.32 ± 0.32 |
| -23.740 | 1.109 | 0.165 | 0.002155 | 6.66 ± 0.54 |
| -22.740 | 1.105 | 0.160 | 0.002146 | 6.18 ± 0.53 |
| -21.736 | 1.101 | 0.155 | 0.002145 | 7.65 ± 0.44 |
| -20.736 | 1.097 | 0.150 | 0.002145 | 7.74 ± 0.34 |
| -19.712 | 1.093 | 0.145 | 0.002136 | 6.65 ± 0.52 |
| -18.712 | 1.090 | 0.141 | 0.002136 | 7.10 ± 0.82 |
| -17.707 | 1.087 | 0.137 | 0.002136 | 7.56 ± 0.35 |
| -16.706 | 1.084 | 0.133 | 0.002135 | 10.49 ± 0.49 |
| -15.684 | 1.081 | 0.129 | 0.002127 | 8.00 ± 0.78 |
| -14.677 | 1.078 | 0.126 | 0.002126 | 8.20 ± 0.27 |
| -13.655 | 1.076 | 0.123 | 0.002126 | 8.04 ± 0.32 |
| -12.655 | 1.073 | 0.120 | 0.002119 | 7.23 ± 0.42 |
| -11.648 | 1.071 | 0.118 | 0.002119 | 8.91 ± 1.46 |
| -10.627 | 1.069 | 0.116 | 0.002119 | 8.70 ± 0.38 |
| 3.704 | 1.060 | 0.134 | 0.002097 | 6.38 ± 0.12 |
| 4.713 | 1.061 | 0.138 | 0.002097 | 7.01 ± 0.45 |
| 5.733 | 1.062 | 0.142 | 0.002097 | 5.93 ± 0.14 |
| 7.742 | 1.064 | 0.150 | 0.002095 | 8.51 ± 0.63 |
| 8.762 | 1.066 | 0.154 | 0.002094 | 5.69 ± 0.33 |
| 9.762 | 1.067 | 0.159 | 0.002093 | 6.74 ± 0.28 |
| 10.771 | 1.069 | 0.164 | 0.002092 | 7.56 ± 0.08 |
| 13.791 | 1.076 | 0.178 | 0.002091 | 7.19 ± 0.12 |
| 14.791 | 1.078 | 0.183 | 0.002090 | 8.02 ± 0.10 |

| ΔT | r | Δ | g | Q ± δQ |
|---|---|---|---|---|
| 15.800 | 1.081 | 0.188 | 0.002090 | 6.12 ± 0.13 |
| 18.650 | 1.090 | 0.203 | 0.002091 | 8.32 ± 0.10 |
| 19.651 | 1.093 | 0.208 | 0.002090 | 8.86 ± 0.12 |
| 20.651 | 1.097 | 0.213 | 0.002091 | 8.75 ± 0.10 |
| 21.660 | 1.101 | 0.219 | 0.002091 | 10.68 ± 0.08 |
| 22.659 | 1.104 | 0.224 | 0.002090 | 11.74 ± 0.08 |
| 23.660 | 1.108 | 0.229 | 0.002092 | 10.24 ± 0.10 |
| 24.659 | 1.112 | 0.235 | 0.002092 | 9.62 ± 0.11 |
| 25.659 | 1.117 | 0.240 | 0.002092 | 8.71 ± 0.10 |
| 26.680 | 1.121 | 0.246 | 0.002091 | 8.82 ± 0.11 |
| 33.688 | 1.156 | 0.284 | 0.002098 | 3.52 ± 0.36 |
| 36.608 | 1.172 | 0.300 | 0.002097 | 3.33 ± 0.17 |
| 37.608 | 1.178 | 0.305 | 0.002105 | 3.70 ± 0.29 |
| 38.608 | 1.184 | 0.311 | 0.002104 | 2.14 ± 0.48 |
| 39.608 | 1.190 | 0.316 | 0.002104 | 2.76 ± 0.52 |
| 40.608 | 1.196 | 0.322 | 0.002104 | 3.53 ± 0.34 |
| 41.608 | 1.203 | 0.327 | 0.002103 | 4.07 ± 0.27 |
| 42.608 | 1.209 | 0.333 | 0.002112 | 2.85 ± 0.35 |
| 43.608 | 1.215 | 0.339 | 0.002112 | 3.02 ± 0.38 |
| 44.607 | 1.222 | 0.344 | 0.002112 | 4.77 ± 0.27 |
| 45.608 | 1.229 | 0.350 | 0.002111 | 6.15 ± 0.23 |

ΔT: Time from perihelion on 2010 October 28.2570 UT in days
r : Heliocentric distance (AU)
Δ: Geocentric distance (AU)
g: Solar Lyman-α g-factor (photons s$^{-1}$) at 1 AU
Q: Water production rates for each image (s$^{-1}$)
δQ: internal 1-sigma uncertainties

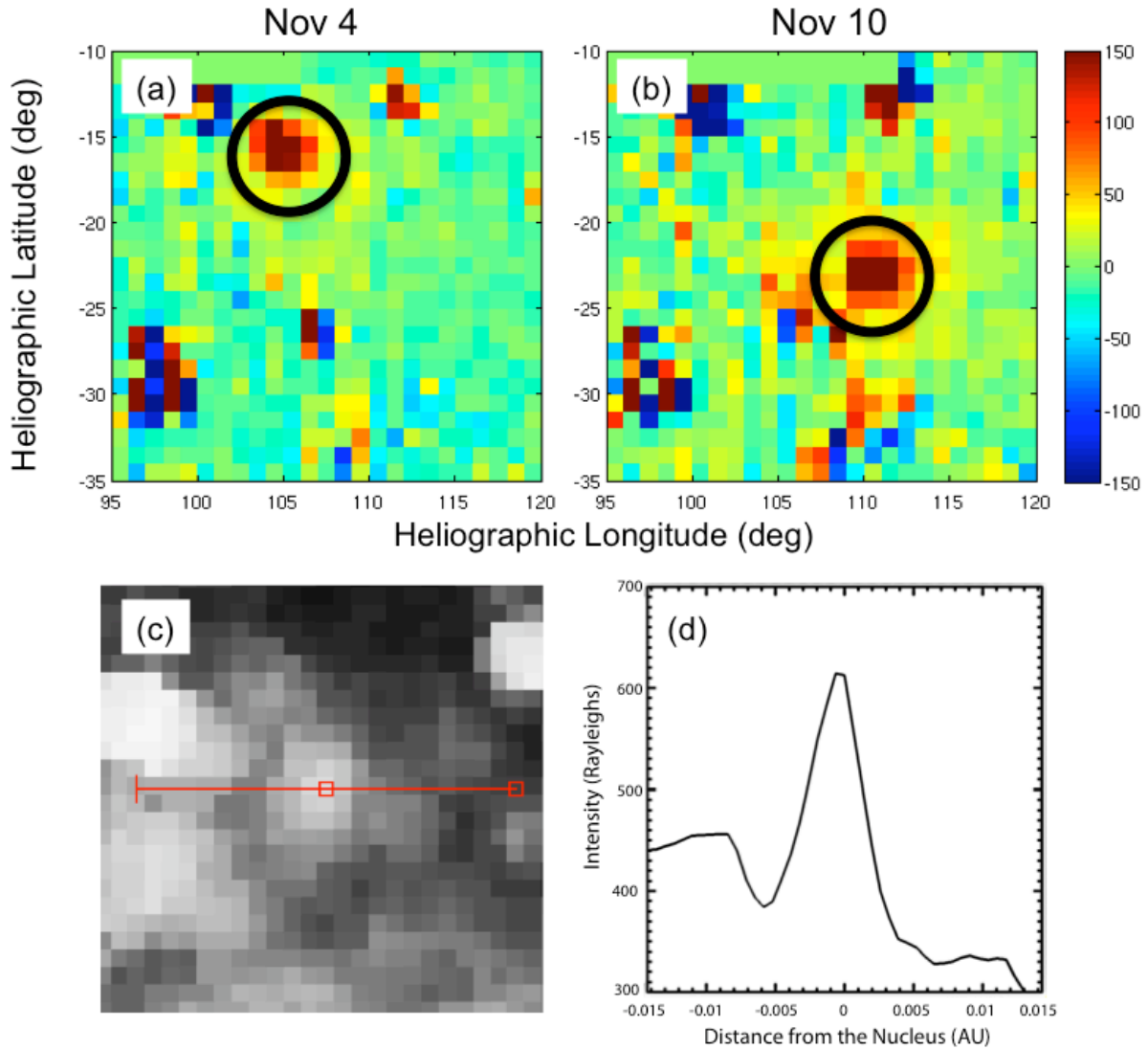

Fig. 1. The sky in H Lyman-α as observed by the SWAN camera on SOHO. In (a) is the sky on 4 November 2010 the day of the EPOXI flyby and in (b) six days later. The comet is highlighted with a black circle. These images are difference images with an image from November 1 subtracted from each. The noise that remains comes from incomplete subtraction of star images owing to inexact spatial registration. The color scale on the right is in Rayleighs. In (c) is an isolated image of the comet from November 4, 2010, projected with the sun to the right, and in (d) is the intensity distribution along the red line cut shown in (c).

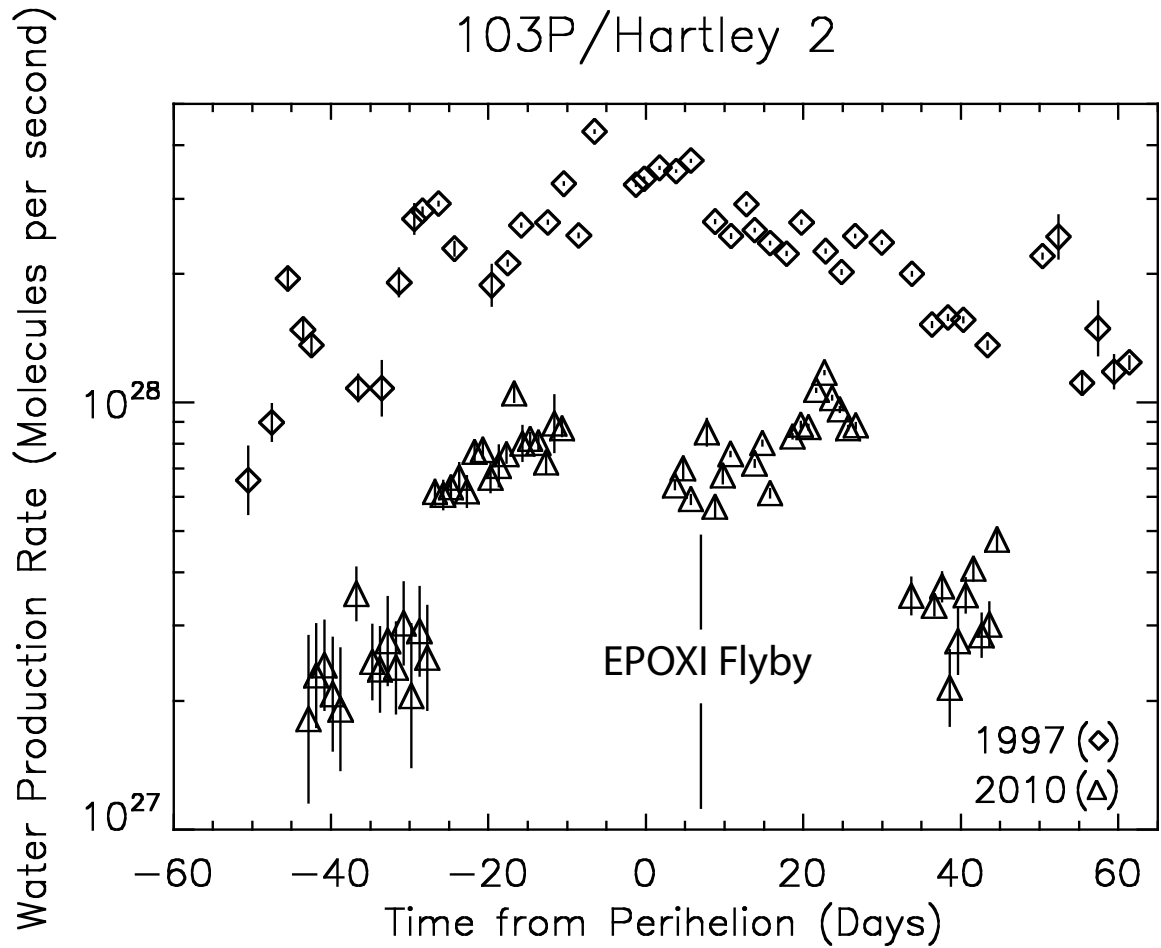

Fig. 2. Water production rate of comet 103P/Hartley 2. The triangles give the water production rates determined from the SWAN H Lyman-α images from September 14 through December 12, 2010. The diamonds give the SWAN results from the 1997 apparition (Combi et al. 2011). The vertical lines give the error bars due to internal sources of error only, namely instrument noise and uncertainty due to the IPM subtraction. The day of the EPOXI flyby in 2010 is indicated. There was a step function increase in water production of about a factor of 2.5 at T=-28 days (October 1). There were data gaps between T=-12 and T=+2 (October 16 to 30) when SOHO spacecraft obstructed the portion of the sky where the comet was located and then again from T=+26 to T=+ 32 (24-30 November) when the comet was too close to nearby stars to get a clean enough image to use. Some time during this period the comet seems to have gone through an activity drop similar to the September 30 rise.

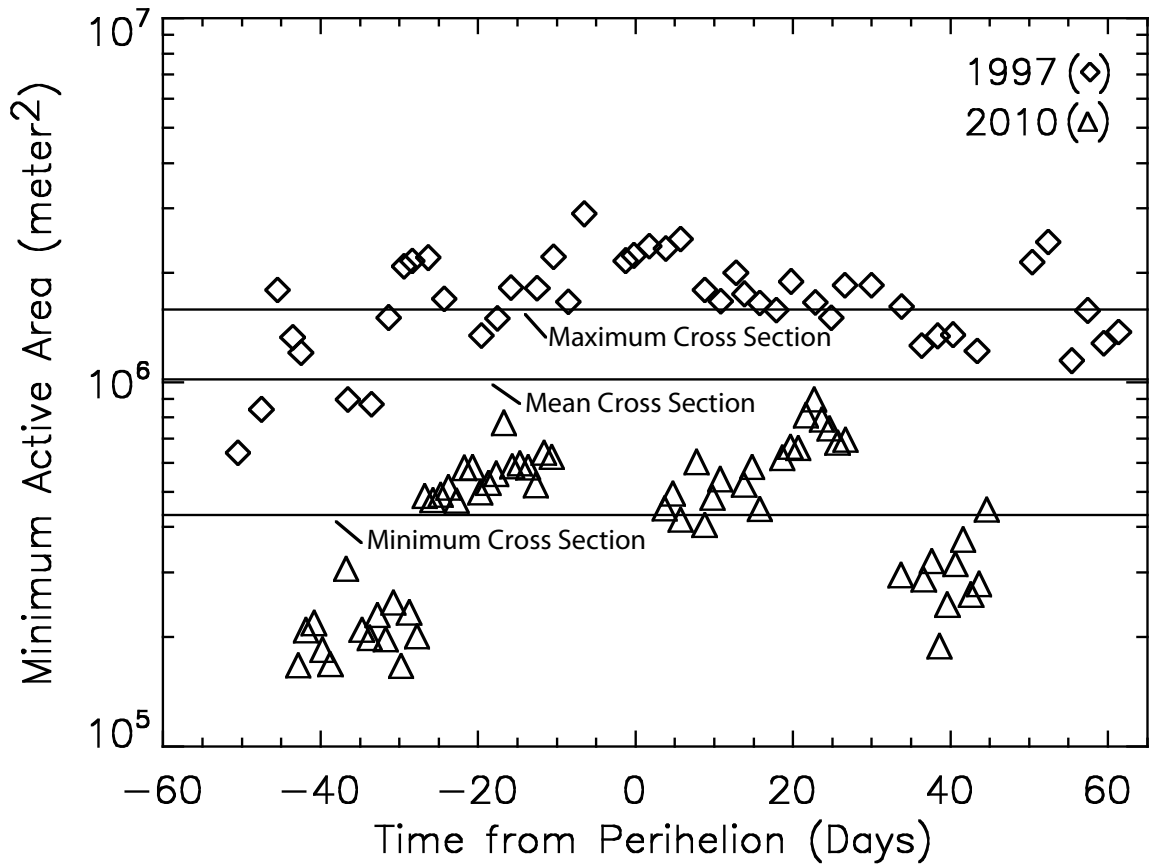

Fig. 3. The Minimum Active Area (in m$^2$) of comet 103P/Hartley 2 plotted as function of time from perihelion. The triangles give the values calculated from the SWAN water production rates in 2010 and the diamonds from 1997. The solid horizontal lines give the maximum and minimum cross sections of the nucleus from EPOXI (A'Hearn et al. 2011) and the mean value from Spitzer Space Telescope (Lisse et al. 2009). Since it is apparent from EPOXI resultsthat the entire nucleus is not active, much of the water production seen within ±20 days of perihelion must be due to the icy fragments released by the $CO_2$-driven activity. Furthermore, since the activity was 3 times higher in 1997, either the $CO_2$-driven activity was much larger, or perhaps some more drastic change happened to the nucleus since 1997.